\documentclass[conference]{IEEEtran}
\IEEEoverridecommandlockouts
\usepackage{cite}
\usepackage{amsmath,amssymb,amsfonts}
\usepackage{tikz}
\usepackage{aeguill}
\usepackage{graphicx}
\usepackage{textcomp}
\usepackage{xcolor}
\usepackage{color}

\usepackage{algpseudocode}
\usepackage{algorithm}

\algnewcommand\algorithmicforeach{\textbf{for each}}
\algdef{S}[FOR]{ForEach}[1]{\algorithmicforeach\ #1\ \algorithmicdo}

\def\BibTeX{{\rm B\kern-.05em{\sc i\kern-.025em b}\kern-.08em
    T\kern-.1667em\lower.7ex\hbox{E}\kern-.125emX}}

\IEEEpubid{\begin{minipage}{\textwidth}
        \centering\footnotesize{\vspace{3.5cm} © 2023 IEEE. Personal use of this material is permitted. Permission from IEEE must be
obtained for all other uses, in any current or future media, including
reprinting/republishing this material for advertising or promotional purposes, creating new collective works, for resale or redistribution to servers or lists, or reuse of any copyrighted
component of this work in other works.}
\end{minipage}} 
\begin{document}
\title{Signaling Storm Detection in IIoT Network based on the Open RAN Architecture
\thanks{This work has been funded by the Polish Ministry of Education and Science within task no. 0312/SBAD/8161, in 2022, and by the National Centre for Research and Development in Poland within project no. CYBERSECIDENT/487845/IV/NCBR/2021}
}

\author{\IEEEauthorblockN{Marcin Hoffmann}
\IEEEauthorblockA{\textit{Rimedo Labs}}
\IEEEauthorblockA{\textit{Institute of Radiocommunications} \\
\textit{Poznan University of Technology}\\
Poznan, Poland
}
\and
\IEEEauthorblockN{Pawel Kryszkiewicz}
\IEEEauthorblockA{\textit{Rimedo Labs}}
\IEEEauthorblockA{\textit{Institute of Radiocommunications} \\
\textit{Poznan University of Technology}\\
Poznan, Poland 
}
}


%


\maketitle

\begin{abstract}
The  Industrial Internet of Things devices due to their low cost and complexity are exposed to being hacked and utilized to attack the network infrastructure causing a so-called Signaling Storm. In this paper, we propose to utilize the Open Radio Access Network (O-RAN) architecture, to monitor the control plane messages in order to detect the activity of adversaries at its early stage.
\end{abstract}


%
\IEEEpeerreviewmaketitle

\section{Introduction}

 One of the significant groups of devices being connected to the 5G and beyond networks is the so-called Industrial Internet of Things (IIoT) devices~\cite{Sisinni2018}. These are mostly stationary deployed sensors used in the industry, e.g., to report the temperature in the furnaces. 
 The main features of IIoT devices are low hardware complexity resulting in low computational power and low cost.  While there can be hundreds of IIoT devices from various vendors in a single network there is a relatively high probability that some of them can be hacked and used to attack the network infrastructure. A possible attack scenario is the so-called Signaling Storm Attack (SSA), where the adversary utilizes standard mechanisms of the network Control Plane (CP) to cause Denial of Service (DoS), e.g., flooding the network CP with invalid or repeated registration requests~\cite{Cao2020}. Even if these registration requests are rejected they consume Core Network (CN) resources in the CP that are needed during the authorization process. It would be beneficial to somehow identify adversary devices in the early stage of the registration process in the Radio Access Network (RAN) to protect CN resources. This can be hard to achieve in state-of-the-art mobile networks, where both hardware and software are provided typically by a single vendor with only a limited possibility of affecting their configuration. On the opposite, the concept of Open RAN (O-RAN) enables interaction with RAN through the dedicated interfaces, and interception of RAN protocol messages~\cite{dryjanski2021toward}. The O-RAN ALLIANCE identified SSA detection as one of the key problems to be resolved by the development of a dedicated xApp~\cite{ORANUseCases}. It should utilize O-RAN interfaces to capture network messages and statistics to detect the abnormal activity of adversaries.

In this paper, we propose a xApp to detect the abnormal activity of IIoT devices at the beginning of their registration procedure. The xApp uses O-RAN interfaces to intercept CP messages to learn the required long-term network statistics. These are used to detect the abnormal activity of adversaries.  



\section{Signaling Storm Detection in O-RAN}

\begin{figure}[!t]
\centering
\includegraphics[width=3.5in]{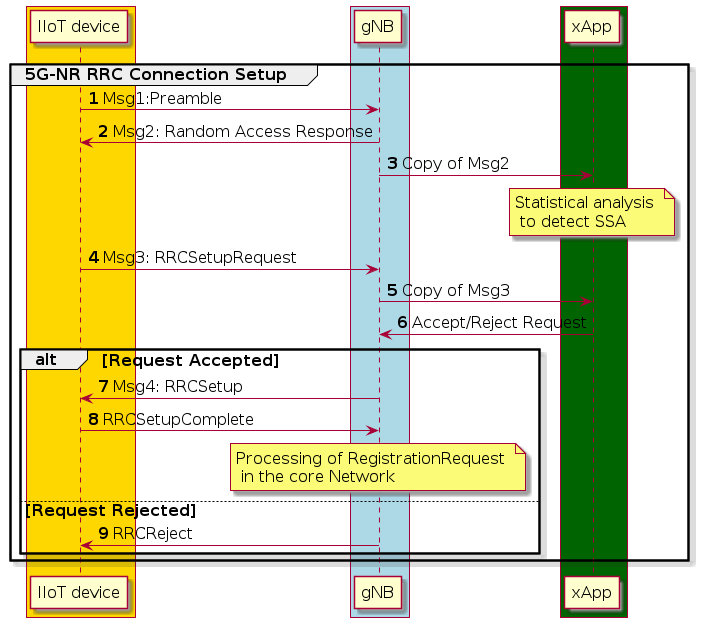}
\caption{CP messages exchanged between IIoT device, gNB, and SSA detection xApp.}
\label{fig:reg_msg}
\end{figure}

\begin{figure}[!t]
\centering
\includegraphics[width=3.3in,height=1.4in]{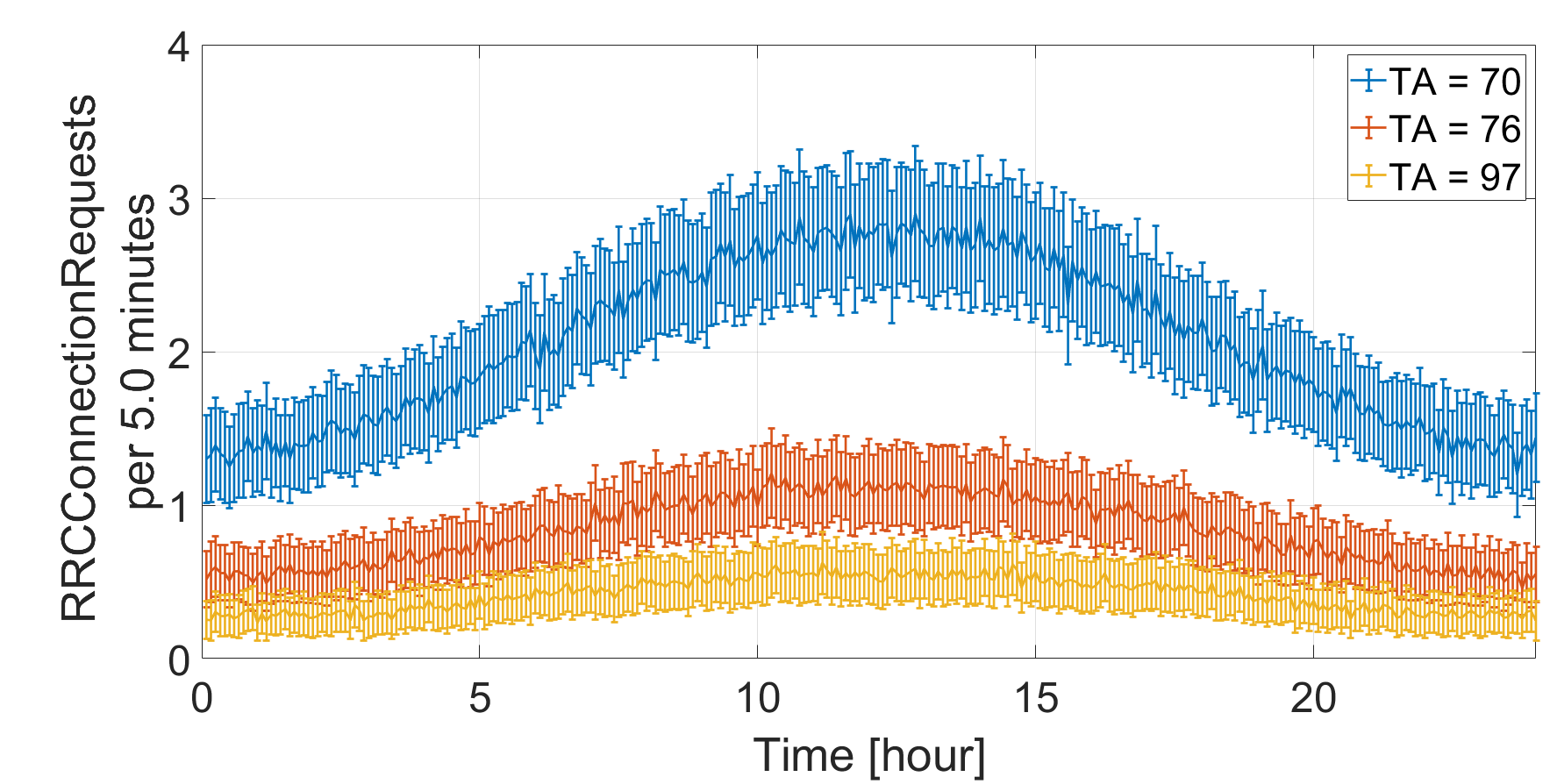}
\caption{Example of KPI profile with a mean and standard deviation of the number of RSRs associated with a given TA per 5 minutes.}
\label{fig:kpi_profile}
\end{figure}

 In order to detect the SSA it is crucial for the xApp to subscribe to messages that are exchanged between the IIoT device and the 5G NodeB (gNB) at the very beginning of the registration procedure. These messages are depicted in Fig.~\ref{fig:reg_msg} based on the 5G New Radio (NR) specification~\cite{3gpp38321}. First, the IIoT device sends Msg1, which contains the random access preamble. After the successful reception of Msg1, the gNB responds to the IIoT device with a Msg2: Random Access Response. 
The Msg2 contains the Timing Advance (TA) command, which is the time offset related to the signal propagation time between the gNB and IIoT devices. As the IIoT devices are in most cases stationary the TA is constant for them and proportional to the distance to the gNB. While the identifier of the IIoT device, e.g., Cell Radio Network Temporary Identifier (C-RNTI) does not have to remain the same between the consecutive registration attempts, during initial access the adversary IIoT devices can be filtered using their specific TA parameters. 
With the use of O-RAN interfaces, the Msg2 can be copied and transferred to the SSA detection xApp for statistical analysis. Following the general framework, from~\cite{bodrog2016}, the so-called Key Performance Indicator (KPI) Profiles, are formulated. The statistics in the KPI Profiles are computed within the consecutive time intervals~$T$, e.g., every 5 minutes. They contain the long-term statistics of RRC Setup Requests (RSRs) associated with observed values of TA, i.e., mean number $\mu_{t,i}$ and standard deviation $\sigma_{t,i}$, where $i$ denotes TA index and $t$ denotes the time of a day period $(t-T;t)$ for which statistics are calculated. An example of such a KPI Profile is depicted in~\ref{fig:kpi_profile}. During the network operation, the number of RSRs is constantly monitored for each observed TA. It is compared against the values from the KPI Profile to compute the so-called anomaly value\cite{bodrog2016}:
\begin{equation}
    a(t,i) = \frac{X(t,i) - \mu_{t,i}}{\sigma_{t,i}},
\end{equation}
where $X(t,i)$ is the number of RSRs observed during time-period $t$ with $i$-th TA. 
If anomaly value $a(t,i)$ exceeds the threshold $\gamma$ 
the SSA is being detected. Based on the detection result the SSA detection xApp indicates the gNB to either accept the RSR and proceed with the registration or reject it. It can be also a general policy send from the xApp, e.g., \emph{Reject all requests of TA=77.}  

\section{Simulation Results}
\begin{figure}[!t]
\centering
\includegraphics[width=3.4in]{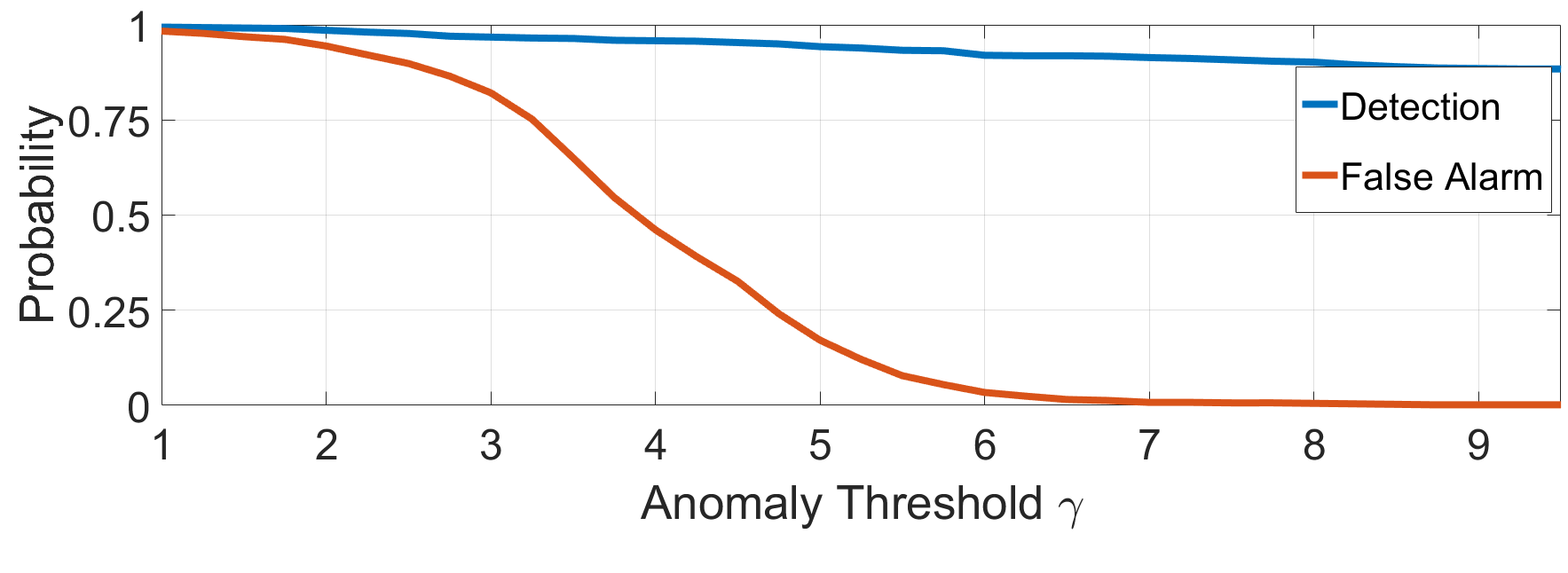}
\caption{Probabilities of SSA detection and false alarm for the proposed xApp as a function of threshold parameter $\gamma$.}
\label{fig:p_det_p_fa}
\end{figure}

The proposed SSA detection algorithm was verified through computer simulations. We have considered a single cell of a 5G-IIoT network of~$2$~km radius with 100 legitimate IIoT devices and 5 adversaries. The legitimate IIoT devices send RSR following the exponential distribution with a mean rate parameter of 5 RSRs per hour. The rate is changing over the daytime resembling a sinusoidal function: it is increased and decreased by $35\%$ at noon, and midnight respectively. Each adversary starts its SSA following the exponential distribution of rate equal to 3 attacks per day. Each attack consists of 100 consecutive RSRs sent within the interval of 5 seconds. In the simulation, we have assumed that the KPI profiles are already computed for the network without adversaries. We have compared the SSA detection for different thresholds of anomaly detection~$\gamma$ for 20-day-long simulations. The resultant probability of false alarm and SSA detection is depicted in Fig.~\ref{fig:p_det_p_fa}. It can be seen that for the low value of $\gamma$ both the probability of detection and the false alarm is very close to $1.0$. The probability of a false alarm can be decreased to $1.5\%$ by setting $\gamma=6.5$ at the cost of the probability of detection degradation to $92\%$. The proposed method allows to balance both probabilities by setting a proper threshold $\gamma$ depending on the network's operator requirements.

\section{Conclusion}
The utilization of O-RAN architecture enables the detection of SSA at an early stage of the adversary registration process. By properly setting the threshold parameter the proposed SSA detection algorithm can offer high detection probability while causing only a few false alarms.

\bibliography{bibliography} 
\bibliographystyle{IEEEtran}

\end{document}